\documentclass{article}
\usepackage[utf8]{inputenc}
\usepackage{setspace}
\usepackage{lineno}
\usepackage{graphicx}
\usepackage{amsmath,amssymb}
\usepackage{makecell}
\usepackage{multirow}
\usepackage{appendix}
\usepackage{adjustbox}
\usepackage{nameref,hyperref}
\usepackage{float}

\title
{Analysis of animal-related electric outages using species distribution models and community science data}

\author
{Mei-Ling E. Feng $^{1+\ast}$, Olukunle O. Owolabi$^{2+}$, Toryn L. J. Schafer$^{3}$,\\
Sanhita Sengupta$^{4}$,  Lan Wang$^{5}$, David S. Matteson$^{3}$,\\ Judy P. Che-Castaldo$^{6}$,
Deborah A. Sunter$^{2,7}$ \\
\normalsize{$^{1}$Conservation \& Science Department, Lincoln Park Zoo, Illinois, USA
}\\
\normalsize{$^{2}$Department of Mechanical Engineering, Tufts University, USA}\\
\normalsize{$^{3}$Department of Statistics and Data Science, Cornell University, USA}\\
\normalsize{$^{4}$School of Statistics, University of Minnesota, USA}\\
\normalsize{$^{5}$Department of Management Science, University of Miami, USA}\\
\normalsize{$^{6}$Branch of Species Status Assessment Science Support, U.S. Fish and Wildlife Service, USA}\\
\normalsize{$^{7}$Department of Civil and Environmental Engineering, Tufts University, USA}\\
\normalsize{$^+$ Authors contributed equally}\\
\normalsize{$^\ast$To whom correspondence should be addressed; E-mail:  mefeng7@gmail.com}\\
}

\date{}

\linenumbers
\begin{document}
\nolinenumbers

\maketitle
\clearpage

\begin{spacing}{2}   

\section*{Abstract}
\begin{enumerate}
    \item Animal-related outages (AROs) are a prevalent form of outages in electrical distribution systems. Animal-infrastructure interactions vary across focal species and regions, underlining the need to study the animal-outage relationship in more species and diverse systems. 
    \item Animal activity has been used as an indicator of reliability in the electrical grid system and to describe temporal patterns in AROs. However, these ARO models have been limited by a lack of available estimates of species activity, instead approximating activity based on seasonal and weather patterns in animal-related outage records and characteristics of broad taxonomic groups, e.g., squirrels.
    \item We highlight publicly available resources to fill the ecological data gap that is limiting joint analyses between ecology and energy sectors. Species distribution models (SDMs), a common technique to model the distribution of a species across geographic space and time, paired with data sourced from eBird, a community science database for bird observations, provided us with species-specific estimates of activity to model spatio-temporal patterns of AROs. These flexible, species-specific estimates can allow future animal-indicators of grid reliability to be investigated in more diverse regions and ecological communities, providing a better understanding of the variation that exists in animal-outage relationship. 
    \item AROs were best modeled by accounting for multiple outage-prone species activity patterns and their unique relationships with seasonality and habitat availability. 
    \item Different species were important for modeling outages in different landscapes and seasons depending on their distribution and migration behavior. 
    \item We recommend that future models of AROs include species-specific activity data that account for the diverse spectrum of spatio-temporal activity patterns that outage-prone animals exhibit. 
\end{enumerate}

\section*{Key Words}
 animal activity; birds; community science; eBird; electrical distribution system; power outages; reliability; species distribution modeling

\clearpage
\section{Introduction}

Power outages result in unwanted disruptions to the functioning of the electric distribution system. These disruptions are a result of a variety of interacting factors including weather, failed equipment, vegetation, and animal activity \cite{maliszewski_environmental_2012}. Understanding system reliability - the ability to withstand disruptions and minimize supply loss - is a long standing research priority to maintain critical infrastructure and essential public services \cite{maliszewski_environmental_2012,gui_advanced_2009}. As a result, assessments of distribution reliability can be broken into three research sectors: historical reliability, predicted reliability, and outage causes \cite{gui_advanced_2009, sekhar_evaluation_2016}. 
Historical reliability is a reactive approach that makes use of historical data on past performance to target less reliable areas of the electric system for improvement. Predictive approaches have developed proactive models that forecast expected outages using a combination of causal factors \cite{gui_bayesian_2011}. The third sector focuses on mechanisms driving outages, identifying the specific factors and their interconnected dependencies that disrupt the electrical system \cite{che-castaldo_critical_2021}. In this study, we focus on the third sector, specifically within the context of animal-related outages (hereafter, AROs), and investigate whether simultaneous activity patterns in multiple species can serve as indicators of reliability in the electric distribution system.

\begin{figure}
\centering
\includegraphics[width=13cm]{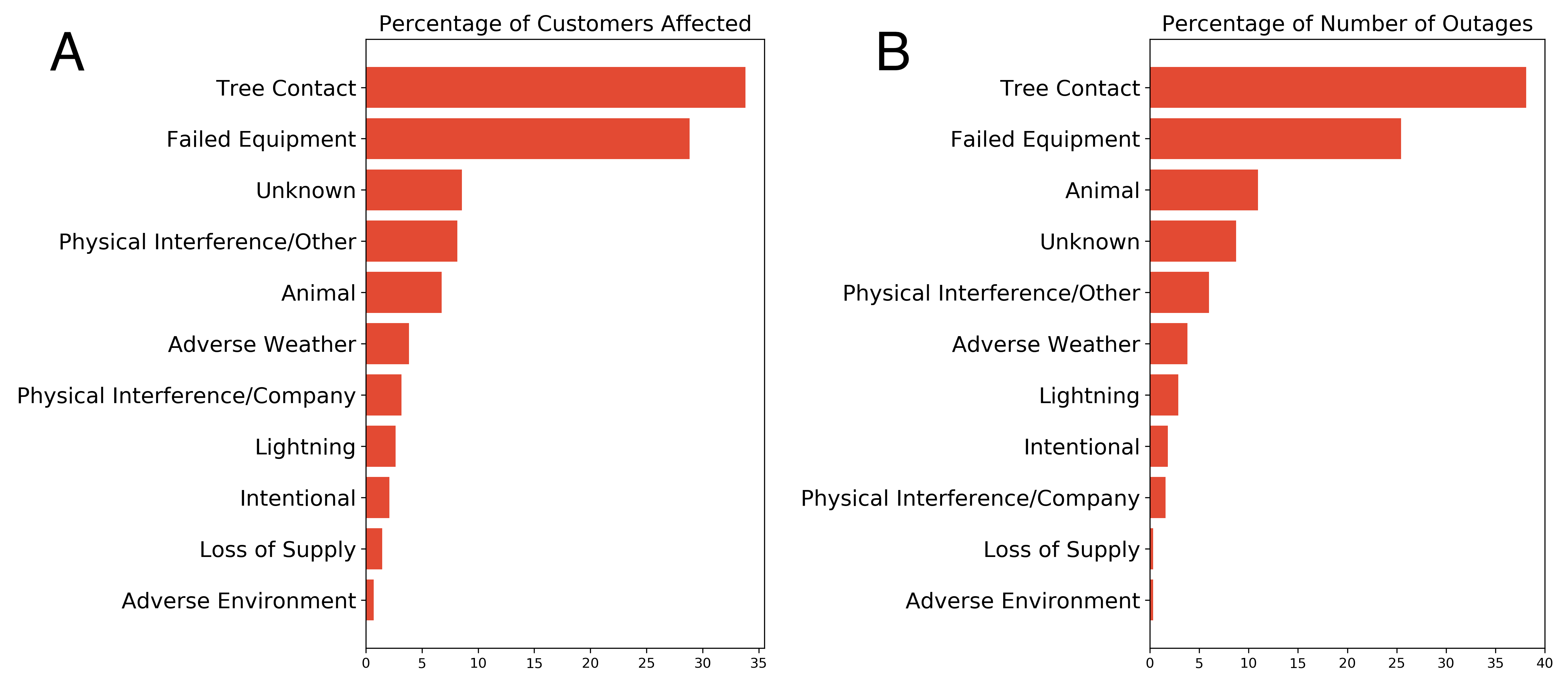}
\caption{(A) Cumulative Percentage of Customers Affected by Causes in MA (2013 - 2018). Initial exploratory analysis showed that AROs accounted for 7\% of total customers affected by electric outages within this period.
(B) Cumulative Percentage of Number of Outages by Causes in MA (2013 - 2018).
 Data obtained from Commonwealth of MA \cite{CommonwealthofMassachusetts2020}.}
\label{fig:rfo}
\end{figure}

AROs contribute a fair proportion of disruptions to the electrical grid. For example, AROs represented about 7\% of the total number of customers affected by electric outages in the state of Massachusetts (hereafter, MA) between 2013  - 2018  (Figure \ref{fig:rfo}). While the number of distribution customers impacted by these AROs are comparatively smaller than other outage causes, they are often frequently occurring, resulting in considerable impact on the grid \cite{Doostan2019}. Many taxonomic groups have been identified as culprits for causing AROs including multiple species of rodents, birds, snakes, ungulates, and medium-sized climbing mammals \cite{frazier_suggested_1996}. The relationship between species populations and grid reliability varies across studies of different species, regions, and habitat types, underlining a need for testing this relationship in more diverse environments and species \cite{loss_refining_2014}. However, a lack of available information on species activity has forced many ARO models to approximate animal activity levels based on seasonal and weather variables derived from life history characteristics in a single taxonomic group, e.g., squirrels \cite{das_outage_2021, gui_bayesian_2011,sahai_probabilistic_2006}. 

Studies in different cities across the United States have shown that non-squirrel taxa such as birds were also important for describing distribution reliability \cite{maliszewski_environmental_2012}. Birds were also listed by the National Rural Electric Cooperative Association (NRECA) as a leading cause of overhead distribution power outages in the United States, second to squirrels \cite{frazier_suggested_1996,avian_power_line_interaction_committee_aplic_suggested_2006}. ARO models that have relied on proxies of animal activity based on a single taxon are limited in their flexibility to be applied to diverse systems. They also lose the ability to account for species-specific mechanisms that drive AROs which allow for more meaningful ARO mitigation strategies \cite{burgio_nest-building_2014,burnham_preventive_2004}. 

Mechanisms driving AROs are derived from environmental conditions and engineering factors that interact with biological species traits \cite{avian_power_line_interaction_committee_aplic_suggested_2006}. Behavioral and morphological traits determine a species' likelihood of interacting with electrical equipment and how this interaction manifests. 

In the case of birds, gregarious social behavior encourages multiple individuals to gather near power lines, causing the lines to sag or swing and collide as flocks take off at once \cite{frazier_suggested_1996,avian_power_line_interaction_committee_aplic_suggested_2006}. Cavity nesting species build nests in electrical equipment and excavate cavities in wooden utility poles. Nesting in equipment increases their chances of collision while flying in and out of the nest, while the nests and cavities themselves damage pole integrity and attract further animals, i.e. predators, to lines \cite{frazier_suggested_1996,avian_power_line_interaction_committee_aplic_suggested_2006,polat_overview_2016}. Perch-hunting strategies make utility poles ideal foraging structures for species of raptors, but also increase their chances of electrocution or collision \cite{avian_power_line_interaction_committee_aplic_suggested_2006}. 

Additionally, morphology such as wing shape, eye placement, and body size can influence the likelihood of collision in flight and electrocution on lines. Wing loading, or the ratio of body size to wing span, can indicate flight maneuverability, while binocular vision is better at detecting obstacles than peripheral vision \cite{damico_bird_2019, bernardino_bird_2018,martin_bird_2010}. Species with larger body sizes and wing spans are able to connect conductors and grounding parts on electrical equipment, causing short circuits \cite{avian_power_line_interaction_committee_aplic_suggested_2006}. 

\begin{figure}
\centering
\includegraphics[width=12cm]{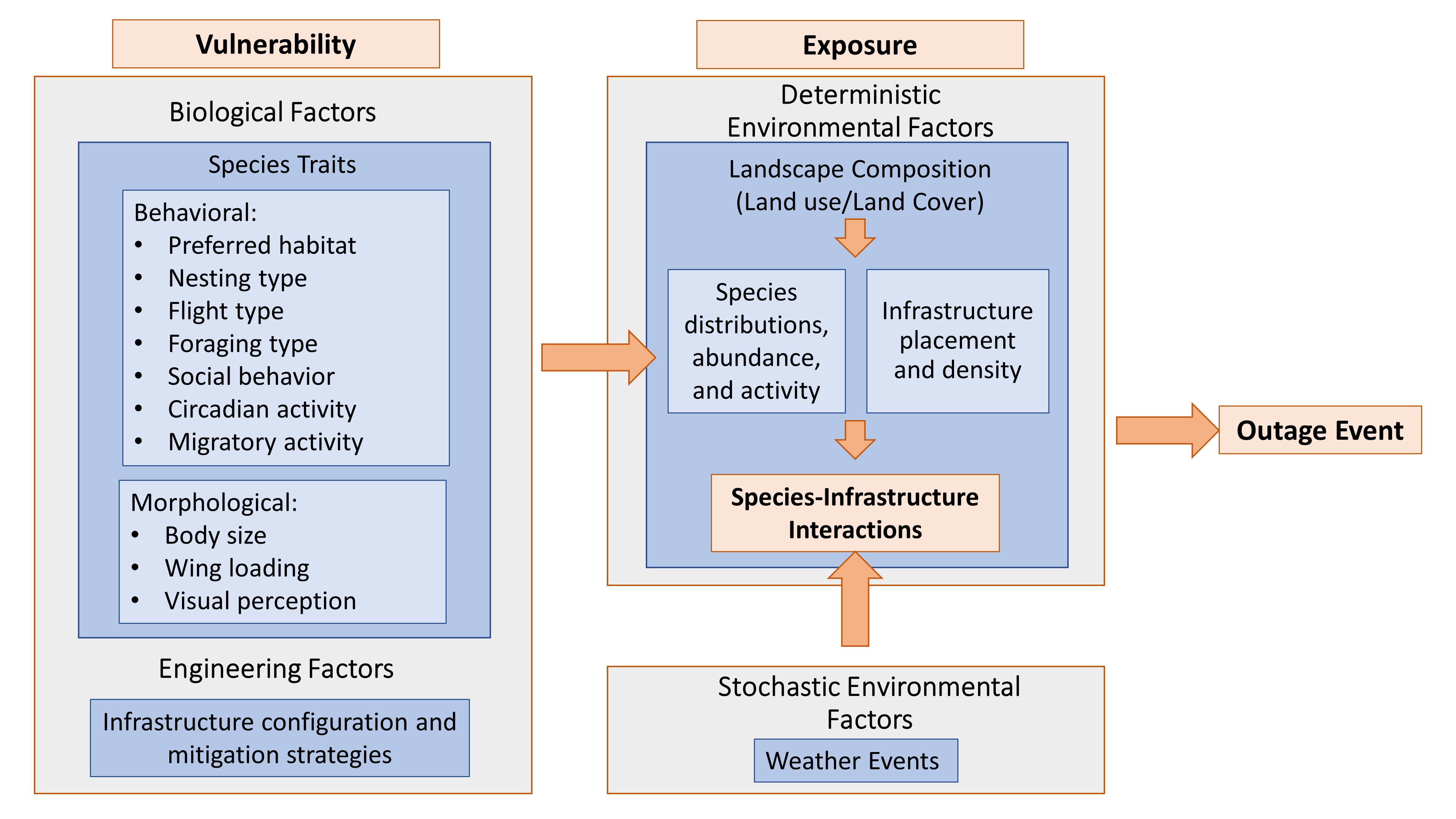}
\caption{Factors contributing to bird related outage risk. Species traits underlie the exposure of vulnerable infrastructure to populations of bird species prone to causing outages.} 
\label{fig:diagram}
\end{figure}

Environmental conditions such as land use and habitat cover and seasonality are important variables that determine the exposure of animals to electrical infrastructure. Different habitats and seasons are suitable to different species depending on their traits and this determines the exposure of species to infrastructure \cite{rollan_modeling_2010,burgio_nest-building_2014,burnham_preventive_2004} (Figure \ref{fig:diagram}). As a result, there is variable ARO risk across space (species distributions) and time (species seasonal activity) which are largely driven by the traits unique to different outage-prone species. This makes considering multiple species important for characterizing the relationship between AROs and ecological patterns.

To derive better indicators of AROs, and simultaneously power grid reliability, more joint analyses of species activity and outage data are needed. The few studies that have used species activity alongside outage data have shown animal activity to be an important variable in characterizing the reliability of the power grid \cite{maliszewski_environmental_2012}, yet these data remain underutilized in animal-outage studies. These joint analyses have been restricted to coarse resolutions and limited species due to the resource intensive nature of biological surveys which limit the accessibility of ecological data \cite{Doostan2019, maliszewski_environmental_2012,sahai_probabilistic_2006}. 

We advocate for the use of community science data to improve the availability of species-specific activity estimates and to provide greater flexibility to explore animal-grid interactions in more diverse systems. Community science databases such as eBird \cite{sullivan_ebird_2009} are comprised of opportunistic species observations submitted by the general public. eBird captures year-round bird observations at a global scale and has best practice methodologies in place to predict fine resolution estimates of species distributions across geographic space and time \cite{strimas-mackey_best_2020,johnston_analytical_2020}. Wide-scale species-specific population data such as eBird is key to better understand and mitigate both bird mortality and ARO disruptions to the electrical distribution system.  

We demonstrate eBird's ability to increase the availability of animal activity for ARO models by using a case study of outage-prone bird species and AROs in the state of MA from 2013-2018. Using eBird data to develop species distribution models (hereafter, SDMs) for 14 outage-prone bird species in MA, we conducted a data-driven exploration of AROs as they relate to spatio-temporal patterns of activity in multiple species. Behavioral and morphological traits differed between our study species, influencing their activity patterns throughout seasons, distributions across the landscape, and their likelihood to cause disruptions to the grid (Figure \ref{fig:diagram}). eBird data allowed us to visualize these patterns in animal activity alongside ARO data at weekly, township resolutions, considered the most accurate temporal resolution at which to estimate AROs \cite{gui_advanced_2009,das_outage_2021}.

Our results show that using species-specific activity improves our ability to model animal-related grid reliability by allowing us to account for multiple spatio-temporal patterns in outage-prone species activity. This multi-species approach is important for understanding ARO electric system reliability, which can vary across space and time depending on how different species are distributed throughout space and time. This approach not only captures bird activity but also habitat characteristics that may explain the occurrence of ARO (including those caused by other animals, i.e., squirrels), making it a promising indicator of animal-related grid reliability. 

Based on these findings, we recommend the use of multi-species activity patterns to improve the accuracy and flexibility of future ARO models. We submit that SDMs based on community science databases are a better and more accessible way to estimate species activity that can then support future joint analyses of ecological and electric distribution systems.

\section{Methods}
All analyses were performed in R statistical language version 4.0.2 \cite{r_core_team_r_2020}

\subsection{Aligning Outage Data and Species Activity Estimates}

\subsubsection{Estimating Bird Activity}
We first identified taxonomic groups of birds known to cause outages due to their behavioral and morphological traits. We select 14 species with these outage prone traits (Table \ref{tab:Table1}) and who also have their year round or breeding range within MA (referenced from eBird species range maps). To ensure we had sufficient detection data, we did not consider nocturnal species such as owls because eBird observations are predominantly recorded during daylight hours. Species that were also not considered in this study but could be considered in future studies in MA are the Monk Parakeet (\textit{Myiopsitia monacbus}), an introduced species increasing in prevalence that builds nests on utility structures, and the American Crow (\textit{Corvus brachyrhynchos}) \cite{avian_power_line_interaction_committee_aplic_suggested_2006}.

\begin{table}[ht]
\centering
\begin{adjustbox}{width=1\textwidth}
\begin{tabular}{|l|l|l|l|}
\hline
\multicolumn{1}{|l|}{Species (Abbreviation)} & 
\multicolumn{1}{|l|}{Disruptive Traits} & 
\multicolumn{1}{l|}{Outage Cause} &
\multicolumn{1}{l|}{References}\\\hline
Brown-headed Cowbird (BHCO) 
& 
\multirow{6}{0.3\linewidth}{Social behavior, nesting type, flight type} &
\multirow{6}{0.3\linewidth}{Phase conductor contact, excrement (contamination), nest contact, predator attraction, collision, electrocution} 
&
\multirow{6}{0.15\linewidth}{\cite{damico_bird_2019,alonso_collision_1999,pettersson_impact_2005,sundararajan_preventive_2004,frazier_suggested_1996}}\\

Common Grackle (COGR)&&&\\
Red-winged Blackbird (RWBL)&&&\\
European Starling (EUST) &&&\\
House Sparrow (HOSP)&&&\\
Mourning Dove (MODO)&&&\\\hline

Downy Woodpecker (DOWO) & 
\multirow{5}{0.3\linewidth}{Nesting type, foraging type}&
\multirow{5}{0.3\linewidth}{Pole damage} & \multirow{5}{0.15\linewidth}{\cite{polat_overview_2016,frazier_suggested_1996}}\\

Hairy Woodpecker (HAWO)  &  & & \\

Northern Flicker (NOFL)  &  &  & 
\\

Pileated Woodpecker (PIWO)  & & & \\

Red-bellied Woodpecker (RBWO) &  &  & \\\hline

Osprey (OSPR)  & 
\multirow{3}{0.3\linewidth}{Nesting type, foraging type, migratory activity, body size} & 
\multirow{3}{0.3\linewidth}{Collision, electrocution, excrement (streamer), nest contact}&
\multirow{3}{0.15\linewidth}{\cite{martin_bird_2010,sundararajan_preventive_2004,frazier_suggested_1996}}\\

Red-tailed Hawk (RTHA)  & & & \\

Turkey Vulture (TUVU)  & & &\\\hline
\end{tabular}
\end{adjustbox}
\caption{\label{tab:Table1}MA bird species selected for their contributions to animal-caused outages. See \cite{feng_comparing_2021} for the list of focal species with scientific names.}
\end{table}

The eBird surveys are semi-structured which allows the data to be standardized and adjusted for common sampling biases. As a result, the data has been used to generate estimates of annual and multi-year trends in species populations consistent with structured bird surveys such as the North American Breeding Bird Survey \cite{walker_using_2017}. Detection data (binary data on whether a species was observed) is more readily available from eBird surveys than individual counts. This makes detection probability (hereafter, DP), the probability of observing a species on a standardized survey, a commonly estimated measure of relative abundance. We use DP as a proxy of activity in birds because animal activity can be an underlying driver of DP, i.e.\ active birds are easier to detect by observers.

To align spatio-temporal resolutions across species' DPs and outage data, we estimate relative bird activity (DP) for our 14 study species at a weekly township-level resolution. We modeled eBird DPs using a random forest approach that appears in an earlier study comparing the performance of multiple modeling methods for relative abundance \cite{feng_comparing_2021}. We apply the random forest approach because it provides added flexibility to account for the multiple spatial and temporal variables needed to characterize species occurrences. The random forest methodology was derived from the "ecounter rate" methodology in the Best Practices for Using eBird Data \cite{strimas-mackey_best_2020,johnston_analytical_2020} which outlines how to estimate species distributions across space and time from the eBird Basic Dataset \cite{noauthor_ebird_2020}. 

The random forest model predicted DP in response to temporal variables (year, day of week, week of the year, and starting time of observations) and survey effort (protocol type and list length). The full data preparation and modeling methodology follows the methods in \cite{feng_comparing_2021}. Here, we add additional spatial predictors in addition to time and survey effort to improve our estimates of species occurrence across both space and time. 

MA township names as well as the proportion of land cover types in each township were also included in the model to account for spatial variation in habitat availability and observer density. Land cover proportions were calculated within each township using 1km resolution land cover types from the National Land Cover Database (NLCD) \cite{HOMER2020184} aggregated into broader habitat types for birds. The final land cover categories used in the RF model were proportion of barren land, cultivated land, developed land, forest, open water, shrub, wetland. Since topography also influences species distributions, the median and standard deviation in elevation within each township were also used as spatial covariates. Elevation data was downloaded from the EarthEnv project \cite{amatulli_suite_2018}. 

\subsubsection{Power Outage Data}
We obtained outage records from three electric distribution utilities (Eversource,  UNITIL, and National Grid) in MA through the MA Office of Energy and Environmental Affairs \cite{DepartmentofPublicUtilities:EnergyandEnvironmentalAffairs}. 
These three utilities
 account for the majority share of electricity distribution in the state. The data range from 2013 - 2018 at a township-level spatial resolution. Fields include the date and time of the outage, township of the outage, and the categorical cause of the outage. The details on our outage data pre-processing appears in the  Supplementary Materials (\nameref{S1_Supplementary_Materials}).




For our analysis, we focused on a subset of outages classified as being caused by an `animal'. We refer to these outages as AROs throughout this paper. AROs were first obtained as a subset of electric outages with a recorded outage cause classified as `Animal', `Animal-other', `Birds', or `Squirrels.'


Previous explorations of animal activity alongside outages have focused on the frequency of ARO events, which is a direct measurement of animal interference with the grid \cite{Doostan2019,maliszewski_environmental_2012,gui_advanced_2009}. However, an important measurement of system reliability in the energy sector is the System Average Interruption Duration Index (SAIDI). SAIDI is used to quantify the amount of time, on average, customers' electricity is disrupted. Although SAIDI incorporates information that can be influenced by other factors outside of animal interactions such as the repair time required by various means of infrastructure design and maintenance strategies \cite{heidari_effects_2013,hanser_control_2015,sekhar_evaluation_2016}, we measure the severity of AROs with SAIDI to test whether animal activity is an indicator of electrical system reliability for grid managers. SAIDI was measured in minutes and is calculated as

\begin{equation}
\label{eq_SAIDI}
    \textrm{SAIDI}_{i, \ell} = \frac{D_{i, \ell}C_{i, \ell}}{H_{\ell}} \, , \quad \textrm{where}
\end{equation}
\\
$D_{i}$ is the duration of the grid disruption in minutes that started at time $i$; \\
$C_{i}$ is the amount of customers affected during the grid disruption that started at time $i$; \\
$H$ is the total number of households served in the system a location $\ell$.
\\\newline

We calculated SAIDI for each outage event before summing SAIDI across all outage records within each township and week to generate our weekly, township-level aggregated outage dataset. To further justify our use of SAIDI over ARO frequency, we tested the relationship between SAIDI and ARO frequency (the number of reported AROs per weekly-township observation) with each of our 14 bird DPs using Pearson's correlation and found that ARO frequency did not show a stronger relationship with bird activity; $r$ values between SAIDI and ARO frequency with DP of each study species was $=< 0.1$).

Our aggregated SAIDI data was positively skewed (mean $=$ 0.99, median $=$ 0.12), so we used the natural log transformation of SAIDI (lnSAIDI) for the remainder of the analysis. We removed outliers of these aggregated lnSAIDI values using the inter quartile range (IQR) method. We took 1.5 times the IQR and added this value to the 75th quantile and subtracted it from the 25th quantile. We then removed week-township records with lnSAIDI 1.5 IQR below the 25th quantile and 1.5 IQR above the 75th quantile. There were several instances of townships with abnormally high weekly values of SAIDI, so this process removed 2.6\% of our week-township outage records, leaving 10,287 observations for the remainder of the analysis.


\subsubsection{Aligning Outage Severity and Bird Activity}
Unlike our modeled species estimates, the outage data was incomplete, meaning not every township or every week had reported AROs. These missing records could be due to unreported cases in addition to true absences, which meant they could only be considered uncertain or `pseudo' absences. Therefore, we only considered the weeks and townships with outage events in our joint analyses of birds and outages. We merged our weekly, township-level outage and bird datasets, dropping 56 townships in the bird dataset that did not have any reported AROs within our study period. Out of 313 weekly observations from 2013-2018 across 351 MA townships (109,863 possible observations), we were left with 10,287 observations that had both species DPs and SAIDI records which accounted for 9.3\% of all possible week-township observations. 
The combined dataset consisted of 15 variables, with lnSAIDI being our measure of ARO severity, and 14 DP estimates, one for each of our study species.  

\subsection{Spatio-temporal Patterns in Outage-Prone Bird Species}

We first applied a Principal Component Analysis (PCA) to examine the spatial and temporal patterns in bird activity (DP) across all species. PCA allows discovery of clear subgroups of species with shared spatio-temporal patterns in their weekly-township observations of DP. We used the \textit{prcomp} R-function to run the PCA on all 14 bird species DPs, accounting for complete weekly observations between 2013-2018 for all 351 MA townships. The resulting DP principal components (PCs) were linear combinations of DPs that explained the most variation in activity across the MA populations of these species. Bi-plots of these principal component loading vectors then group species with distinct spatial and temporal patterns of activity.

We grouped species with similar spatio-temporal variation determined by the first two PCs, i.e. species with positive PC values, species with negative PC values, and species with values close to zero (Figure \ref{fig:biplots}), and plotted their state average DPs to visualize distinct temporal activity patterns in our study species. We also looked at their distinct spatial patterns by mapping mean species DP averaged across time for each MA township.


\subsection{Accounting for Several Activity Patterns in Models of Grid Reliability}
We compared the performance of multiple linear regression models using different combinations of species, habitat, and temporal predictors to assess whether 
(1) SAIDI is best predicted using animal activity or through its own historical seasonal and spatial variation, 
(2) SAIDI is better explained by multiple activity patterns or a single pattern of species activity, 
and 
(3) whether the importance of certain species for predicting SAIDI changes in different landscapes and seasons of the year. 

In each model we used lnSAIDI as our dependent response variable. 
For our predictors, we used species DPs to represent animal activity, the proportion of land cover occupied by distinct habitat types in each township (see Section 2.1.2), and temporal variables of month and year as factors. To reduce collinearity between predictors, we removed the developed land habitat type because it was highly anti-correlated with forested habitat and retained only one representative blackbird species (Red-winged blackbird) and three representative woodpecker species (Red-bellied woodpecker, Hairy woodpecker, and Pileated woodpecker) because DP dynamics between species in these families were highly correlated. This left 10 of our 14 bird species which were used for the subsequent SAIDI models.

To assess whether animal activity improved our models of SAIDI, we compared seven different baseline models. One model used only habitat types (proportion of forest, grassland, barren land, and open water) as explanatory variables, another used only time variables, and a third used only
species detection probabilities for our 10 remaining bird species. 
We also include a model using
time and habitat variables to infer whether animal caused SAIDI could be simply explained using its own historical seasonal and spatial patterns independent of animal activity. 

We next considered models using
species and time, and species and habitat, to infer whether seasonal or spatial patterns in species activity were more important indicators of SAIDI. In these models we included interaction effects between species DP and month, and DP and forest cover (a proxy for natural habitat). Previous studies have shown animals are more likely to use power line structures for perching, roosting, and nesting in the absence of natural structures and habitat \cite{avian_power_line_interaction_committee_aplic_suggested_2006}. However, an alternative hypothesis is that species are likely to be more active in areas with suitable habitat, also increasing ARO risk. 
Our final model included species, habitat, and time based on statistically significant (p-value$<$0.05) interaction terms from previous models. We compared model performance using AIC, R squared values, and residual standard error. 

To assess which animal activity patterns where the most important indicators SAIDI, we then compared models using all 10 study species against models using only species that exhibit the same spatial and temporal patterns based on their loading vectors from the PCA. For this model comparison we tested a model using the interactions of species, habitat, and time using all study species, only species with distributions concentrated in more forested, rural areas of MA, only species concentrated in more developed, urban areas of MA, only summer migrants, and only year-long resident species. We repeated the same model comparison of these `All species', `Urban species', `Rural species', `Migrant species', and `Resident species' models to see whether accounting for multiple patterns of species activity outperformed models only accounting for single patterns. 

We expect that multiple species activity patterns are important for describing AROs through time across a region such as MA. We hypothesize this because species with different behavioral traits are distributed unevenly across a landscape and through seasons, exposing power lines to different sets of outage-prone species at smaller resolution, distinct geographic locations and times of year. As a result, we expect important species contributors to AROs to change in different seasons and landscapes, depending on their exposure to power lines through their activity patterns. 

To assess whether certain species better capture changes in SAIDI in different seasons and landscapes, we applied our SAIDI model using all study species to temporal and spatial subsets of our SAIDI data. Our two temporal subsets included only weeks during the summer months, which we defined as June through August, and weeks only during the winter months (December through February). Our two spatial subsets looked at observations only in townships with the highest amount of natural habitat ($>$50\% forest coverage), and with the highest amount of urban habitat ($>$50\% developed land coverage). For the temporal subsets we included only habitat interactions with species DP and for the spatial subsets we included only monthly interactions with species DP. We compared species coefficients and their significance (p-values) in each of these models `Summer', `Winter', `Forested', `Urban') to see if each species' significance and relationship with SAIDI changed during different times of year and in different landscapes. 

\section{Results and Findings}

\subsection{Spatio-Temporal Patterns in Bird Activity}
\begin{figure}[htp]
\centering
\includegraphics[width=13cm]{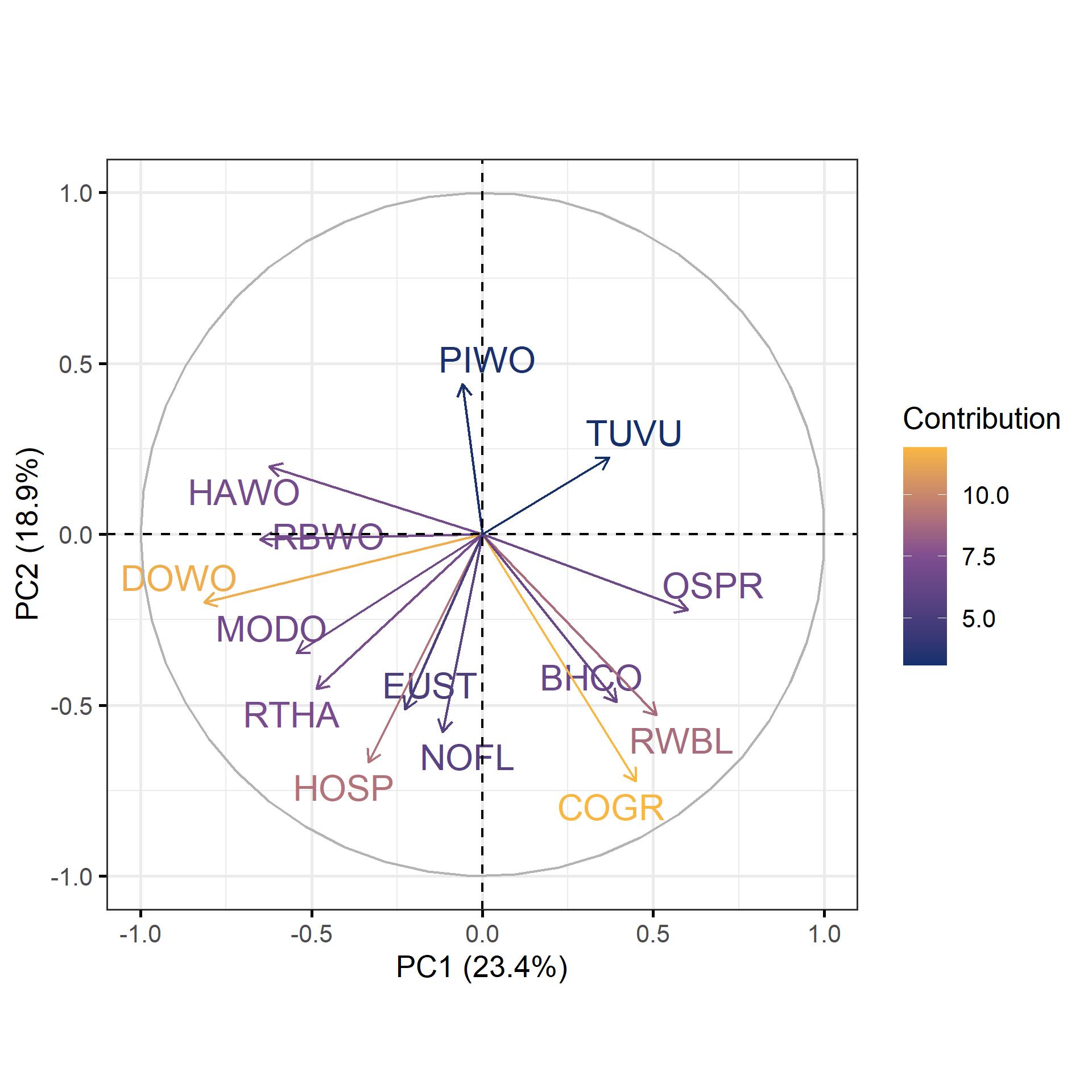}
\caption{The first two principal components (PCs) of weekly detection probabilities across 14 outage-prone bird species in MA townships. Each species is placed along each axis by its loading vector in the first PC and the second PC. These two axes divide species into their unique patterns in activity throughout seasons and spatial distributions across the landscape. The color and length of each vector indicates each species' contribution to the variance in the PCs. Species in the bottom right quadrant are summer migrants limited to urban environments, while the upper left quadrant are resident species occupying more rural areas of the state. The upper right quadrant contains migratory species occupying rural regions and the bottom left quadrant contains resident species concentrated in urban regions.} 
\label{fig:biplots}
\end{figure}

\begin{figure}[H]
\centering
\includegraphics[width=13cm]{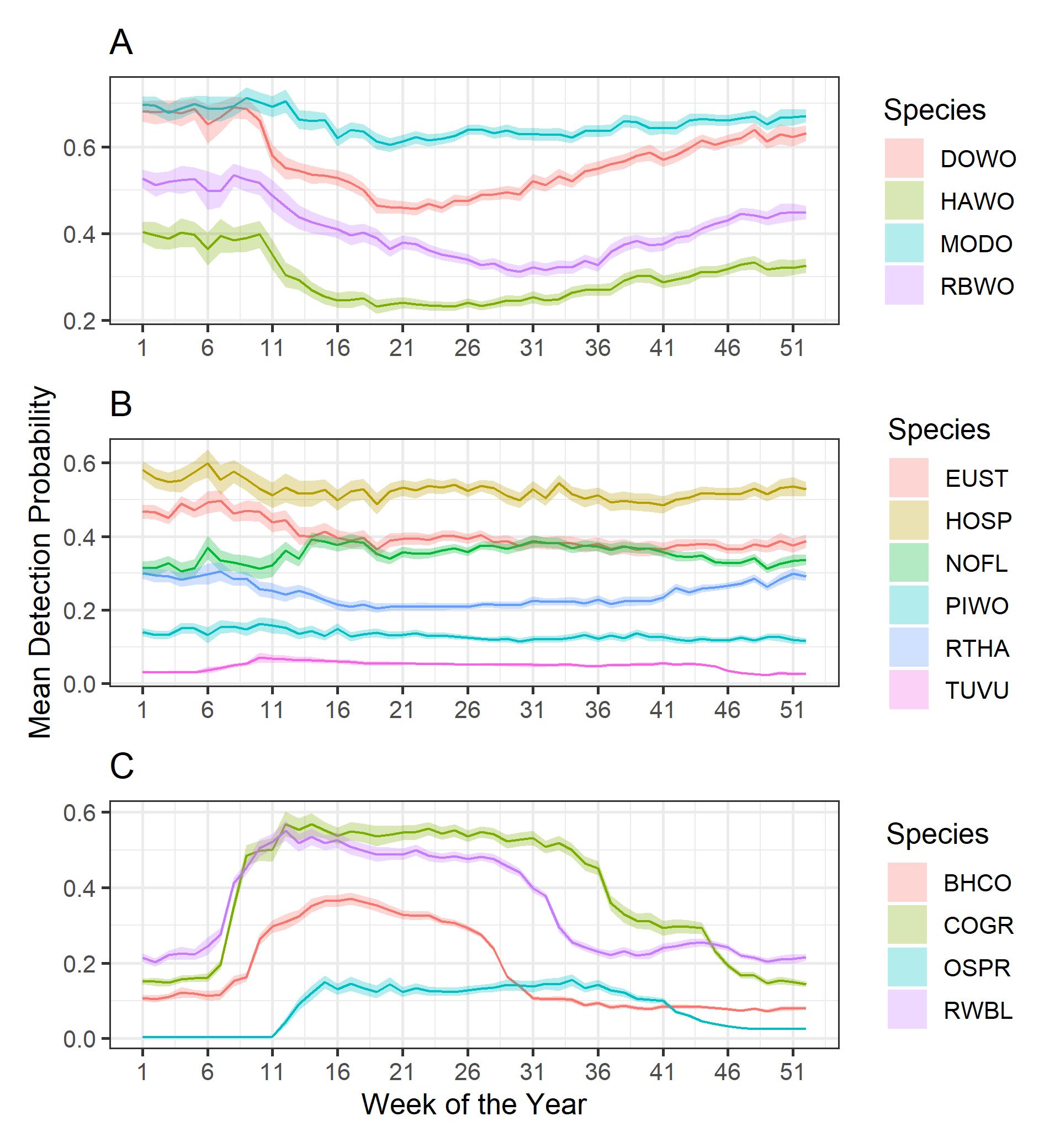}
\caption{Mean and 95th percentiles of species detection probabilities, averaged across years and townships, are plotted for each week of the year. Resident species in group A) have higher winter activity, species in B) did not express strong seasonal activity patterns, and migratory species in C) have higher summer activity in MA. These species were grouped by their PC1 loading vector values.} 
\label{fig:ts}
\end{figure}

The first two PCs of our bird DP PCA explained 40\% of the variance across species DPs and highlighted the main spatio-temporal patterns of activity seen in our 14 focal species (Figure \ref{fig:biplots}). By visualizing time series and maps of DPs for species grouped by their loading vectors along the first two PCs, we found the first PC largely grouped species based on their seasonal activity patterns, i.e.\ heightened summer activity in migratory species such as Red-winged blackbird (RWBL) (Figure \ref{fig:ts}, C) versus heightened winter activity in resident species such as Red-bellied woodpecker (RBWO) (Figure \ref{fig:ts}, A). 

\begin{figure}[H]
\centering
\includegraphics[width=13cm]{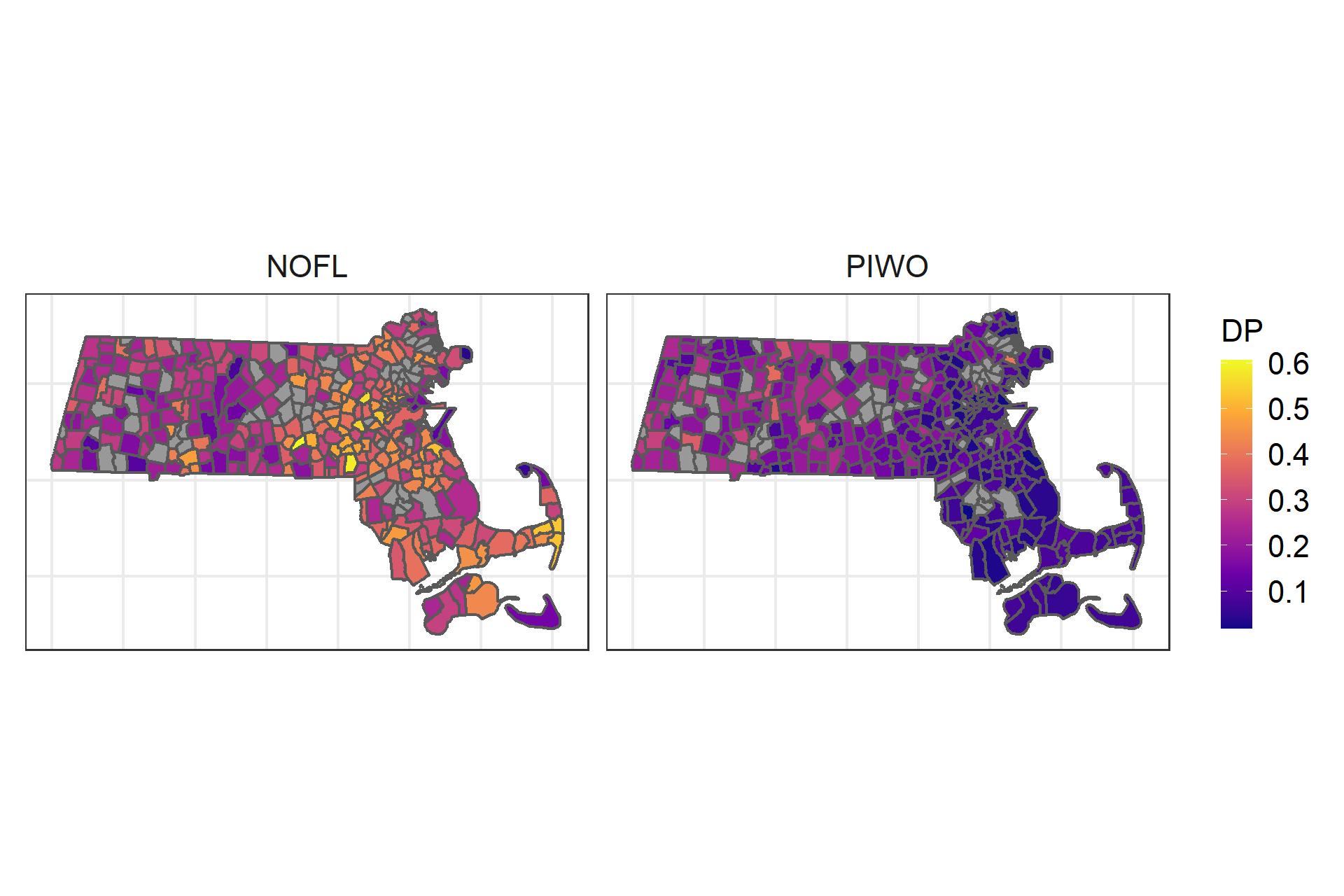}
\caption{Maps of the average weekly detection probabilities for Pileated Woodpecker (PIWO) and Northern Flicker (NOFL) in MA townships. These species display two distinct spatial distributions across the state. PIWO are more frequently observed in continuous forested habitat, while NOFL can occupy forest patches in more urbanized environments surrounding the cities of Springfield and Boston. See our Supplementary Materials for maps of all 14 study species (\nameref{S2_Supplementary_Materials}).} 
\label{fig:maps}
\end{figure}

Given the first grouping, the second PC further grouped species based on their spatial distributions, i.e.\ Pileated woodpecker (PIWO) which occupies continuous forested areas versus Northern flicker (NOFL) which can occupy forest patches in more urbanized environments (Figure \ref{fig:maps}). See our Supplementary Materials for maps of all study species' distributions and land cover types in Massachusetts (\nameref{S2_Supplementary_Materials}). Together these PCs indicate four distinct spatio-temporal patterns of bird activity, i.e.\ rural resident species, urban resident species, rural migratory species, and urban migratory species. Plotting species by their loading vectors along these two PCs grouped our study species into these four distinct activity patterns (Figure \ref{fig:biplots}). 

\subsection{Grid Reliability vs Species Activity Patterns}
Our first model comparison tested whether animal-caused SAIDI is best modeled using animal activity or by its historical seasonal and spatial patterns. We find that the best performing model included species activity, habitat, and time variables (Table \ref{tab:Table2}). Accounting for species activity and their unique interactions with habitat and seasonality accounted for the most variation in SAIDI (R squared $=$ 0.07) and had the lowest residual standard error and delta AIC.

\begin{table}[ht]
\centering
\begin{tabular}{|l|l|l|l|}
\hline
\multicolumn{1}{|l|}{SAIDI Model} & 
\multicolumn{1}{l|}{RSE} &
\multicolumn{1}{l|}{R Squared}&
\multicolumn{1}{l|}{Delta AIC}\\\hline
Species * Time + Species * Habitat&
1.41&
0.07&
0.00
\\\hline
Species * Habitat &
1.43&
0.05&
81.45\\\hline
Time + Habitat &
1.43&
0.05&
102.92\\\hline
Habitat &
1.44&
0.04&
234.67\\\hline
Species * Time &
1.43&
0.06&
253.75\\\hline
Species&
1.44&
0.03&
299.88\\\hline
Time &
1.46&
0.02&
495.49\\\hline

\end{tabular}
\caption{\label{tab:Table2} Models of SAIDI improved when accounting for observed species activity in addition to its underlying seasonal and spatial patterns. We compared seven multiple linear regression models with different combinations of explanatory variables of animal-related SAIDI. All models had f-tests in comparison to an intercept only model that were significant at the 0.1\% level (p-value$<$0.01). Accounting for the interactions between species activity, habitat availability, and months within the year out performed models that only accounted for part of these interactions or explanatory variables. * indicates an interaction term.}
\end{table}

Next we asked whether including species that represented diverse spatio-temporal patterns in activity improved SAIDI model performance or if SAIDI was mostly dependent on single pattern of animal activity specific to certain species. Modeling SAIDI using all 10 study species, representing diverse spatial distributions and seasonal trends in activity, outperformed models that only used subsets of species that exhibited similar patterns in activity as determined by PCA (Table \ref{tab:Table3}).

\begin{table}[ht]
\centering
\begin{tabular}{|l|l|l|l|}
\hline
\multicolumn{1}{|l|}{SAIDI Model} & 
\multicolumn{1}{l|}{RSE} &
\multicolumn{1}{l|}{R Squared}&
\multicolumn{1}{l|}{Delta AIC}\\\hline
All Species &
1.42&
0.07&
0.00\\\hline
Migrants &
1.43&
0.05&
106.89\\\hline
Rural &
1.43&
0.04&
181.94\\\hline
Urban &
1.44&
0.03&
318.91\\\hline
Residents &
1.45&
0.02&
455.71\\\hline

\end{tabular}
\caption{\label{tab:Table3} A comparison of multiple linear regression models used to select the most important patterns in species activity for modeling animal-related SAIDI. All models had f-tests in comparison to an intercept only model that were significant at the 0.1\% level (p-value$<$0.01). Accounting for multiple activity patterns unique to different groups of species out performed models that only accounted for a single spatio-temporal pattern (e.g. summer migrants versus year-long resident species and species occupying urban versus mature forest habitat). Models of AROs can benefit from incorporating activity data from multiple diverse species.}
\end{table}

In both sets of model comparisons, the model accounting for individual species activity across diverse species in addition to their interactions with habitat and time had the best performance. We applied this best performing model to spatial and temporal subsets of our SAIDI dataset and found species importance changed when accounting for variation SAIDI in different landscapes and seasons. When accounting for all weeks in the year and townships across the state, increases in European starling (EUST) activity had the only significant relationship with increases in SAIDI or outage severity. European starling remained a significant positive contributor to outages across seasons and was the only significant contributor to outages in the winter season, while Red-winged blackbird becomes an additional significant contributor to outages during the summer months. European starling and Red-winged blackbird continue to be significant outage contributors in urbanized areas, but are replaced by Turkey vulture (TUVU) and Red-tailed hawk (RTHA) as the significant contributors to outages in heavily forested areas (Table \ref{tab:Table4}).

\begin{table}[ht]
\centering
\begin{tabular}{|l|l|l|l|l|l|}
\hline
\multicolumn{1}{|l|}{Species} & 
\multicolumn{1}{|l|}{All Space and Time} & 
\multicolumn{1}{l|}{Summer} &
\multicolumn{1}{l|}{Winter}&
\multicolumn{1}{l|}{Forested}&
\multicolumn{1}{l|}{Developed}\\\hline
TUVU &
0.09& 
-1.02&
1.07&
\textbf{2.51*}&
-3.38\\\hline
MODO &
0.65& 
0.68&
0.74&
0.66&
2.86\\\hline
HOSP &
\textbf{-1.26*}& 
\textbf{-0.43*}&
\textbf{-0.87*}&
-1.68&
\textbf{-2.48*}\\\hline
OSPR &
-0.28& 
\textbf{-0.58*}&
0.02&
0.75&
\textbf{-1.76*}\\\hline
RTHA &
\textbf{-2.02*}& 
\textbf{-2.88*}&
\textbf{-2.27*}&
\textbf{1.49*}&
-0.46\\\hline
RWBL &
0.80& 
\textbf{0.87*}&
0.33&
-3.61&
\textbf{4.74*}\\\hline
PIWO &
\textbf{-2.18*}& 
\textbf{-3.20*}&
-0.99&
0.26&
-0.95\\\hline
HAWO &
0.82& 
0.49&
0.67&
-0.88&
1.29\\\hline
NOFL &
\textbf{-0.75*}& 
\textbf{-0.68*}&
\textbf{-1.32*}&
\textbf{-1.18*}&
-0.82\\\hline
EUST &
\textbf{1.31*}& 
\textbf{1.85*}&
\textbf{1.61*}&
0.66&
\textbf{0.98*}\\\hline
\end{tabular}
\caption{\label{tab:Table4}Different species are important for understanding animal-caused SAIDI changes in different landscapes and seasons. A multiple linear regression model of SAIDI using relative activity of 10 outage-prone bird species and their interactions with habitat and months was applied to subsets of SAIDI records. These subsets included only records during the summer months, winter months, in heavily urbanized MA townships, and heavily forested MA townships. The coefficients of each species relative activity and their significance in each data subset is reported. * indicates a p-value $<$ 0.05}
\end{table}


\section{Discussion}
\subsection{Improving ARO models using Diverse Animal Activity Patterns}
Outage-prone bird species in MA exhibit diverse spatio-temporal patterns in activity due to their varying behavioral traits and life histories. Migratory species such as Osprey (OSPR) and Red-winged blackbird have the strongest seasonal fluctuations and are nearly inactive in the state during the winter months (Figure \ref{fig:ts}, C), while late spring into fall have the highest levels of activity during breeding and migration. Resident species such as House sparrow (HOSP) and Downy woodpecker (DOWO) have more stable year-long activity and increase in activity during the winter likely due to northern populations moving south and overwintering in the state (Figure \ref{fig:ts}, A).
Species such as Pileated Woodpecker, who occupy large, intact mature forests are distributed across forested areas in rural sections of MA (Figure \ref{fig:maps}). In contrast, opportunistic species such as the Northern Flicker can take advantage of smaller forest patches within the suburbs of urbanized areas in the state. Our first PC detected these unique seasonal patterns while the second PC, given the first, describes these distinct rural versus urban spatial distributions in outage-prone species.

These diverse patterns in bird activity result in varying degrees of species exposure to MA' electric distribution system. By applying our SAIDI model that accounted for multiple species activity patterns to different spatio-temporal subsets, we find that different outage-prone species are significant contributors to outages across different seasons and landscapes. This is likely due to the differences in activity patterns determined by their behavioral traits. For instance, resident species such as European starling occupy MA year-round and are a constant threat to equipment across seasons, while migratory species such as Red-winged blackbird are mostly exposed to equipment from spring through fall. Spatial distributions also determine the exposure of species to electrical equipment. Turkey vultures and Red-tailed hawks become an important contributor to AROs in heavily forested areas, while urban, generalist species such as European starling become more important contributors to AROs in developed areas (Table \ref{tab:Table4}). 

Previous studies of AROs have linked temporal patterns in AROs to the behavioral characteristics of a single taxonomic group, e.g.\ squirrels, in order to understand disruptions of the electric distribution system \cite{gui_advanced_2009,gui_bayesian_2011,sahai_probabilistic_2006,Doostan2019}. To demonstrate the drawbacks of relying on a single animal activity pattern, we found ARO models using species activity that represented diverse patterns across seasons and the landscape out-performed models that used species representing a single spatio-temporal pattern (Table \ref{tab:Table3}). Our findings indicate that multiple spatio-temporal patterns in animal behavior can contribute to AROs and serve as a more holistic indicator of grid reliability at a regional scale. 

\subsection{Data Considerations for Future Studies of AROs}
Previous studies have found broader groups of species contribute differently to AROs in different regions \cite{frazier_suggested_1996}. For instance, occurrences of AROs in Kansas have been mostly closely tied to activity patterns found in squirrels \cite{gui_advanced_2009}, while birds were shown to be an important predictor of outages in Arizona \cite{maliszewski_environmental_2012}. In our study, we found that birds have a low, but detectable effect on AROs, with the best model of AROs using bird activity only explaining 7\% of the variance in SAIDI (Table \ref{tab:Table2}). It could be that another taxonomic group of wildlife such as rodents play a larger role in causing AROs in MA. Better data is need to begin incorporating activity data for more diverse taxa, especially for small mammals. Existing species occurrence databases such as i-Naturalist and the Global Biodiversity Information Facility (GBIF) \cite{noauthor_gbif_nodate} do not compile complete survey data. As a result, they cannot account for sampling bias and true absences, a benefit that allows eBird to produce more accurate SDMs.

Animal activity is not the only factor that drives animal-infrastructure interactions. In some cases it is the increasing susceptibility of existing populations that plays a greater role in determining grid interactions. Stochastic variables such as weather events can further drive bird-infrastructure interactions by increasing bird susceptibility to line collisions, while data on infrastructure composition can improve knowledge of species-infrastructure overlap and exposure. In these cases, the spatio-temporal distribution of species may be important for explaining animal activity, while species distributions in conjunction with stochastic environmental data such as extreme weather events and information on infrastructure may add further understanding to the relationship between animals and electrical grid reliability (Figure \ref{fig:diagram}). Therefore, the use of animal activity as an indicator of grid reliability could be strengthened by including additional environmental and infrastructure data alongside estimates of species activity. 

A further data limitation in our study was the inconsistent reporting in the outage data we obtained from utility companies. The reported ARO causes in our data consisted were mostly non-descriptive classifications, i.e. `animal' or `animal-other,' with taxon specific causes identifying between birds and squirrels limited to a few hundred records. In many cases AROs can be difficult to report, with animals moving away from the site after causing an outage, or predators carrying away carcasses \cite{avian_power_line_interaction_committee_aplic_suggested_2006}. Legal processes can also deter reporting of species protected by conservation laws. As a result, accurate ARO data and their specific causes can be difficult to extract from reported outage records. In order to improve our understanding of AROs, there continues to be a need for better data in both the ecology and energy sectors.


\section{Conclusions}
Making the best available resources from quantitative ecology known to outside disciplines can allow for more successful interdisciplinary data analyses between ecological and non-ecological disciplines. Species distribution modeling is used by ecologists to estimate the distribution of species across continuous spatio-temporal resolutions from observational data collected by limited biological surveys. We suggest that SDMs in combination with more widely available community science databases can be used to improve the availability of animal activity estimates in models of AROs and allow these models to be applied to more diverse regions.

Species contributing to AROs vary across regions due to variation in environmental conditions, species composition, and infrastructure planning. By identifying diverse spatio-temporal activity patterns for outage-prone bird species in the state of MA, we show that electric system reliability across a broad region was best described by using a combination of seasonal activity patterns and spatial distributions unique to different species. We recommend considering and using multiple patterns of species activity as indicators of grid reliability since this can improve our ability to describe the variation in animal-grid interactions through seasons and also within and across geographic regions. 


\section*{Supporting information}

\paragraph*{S1 Supplementary Material}
\label{S1_Supplementary_Materials}
{Outage data preprocessing methods.} 

\paragraph*{S2 Supplementary Material}
\label{S2_Supplementary_Materials}
{Maps of habitat coverage and average detection probability for 14 outage-prone bird species across Massachusetts townships.}

\section*{Acknowledgments}
This work was made possible by the MA office of Energy and Environmental Affairs \cite{DepartmentofPublicUtilities:EnergyandEnvironmentalAffairs}, through which we obtained outage datasets for Eversource Energy, National Grid, and Unitil Corporation as used in this work, as well as the Cornell Lab of Ornithology for their open source bird data \cite{sullivan_ebird_2009}. We thank the volunteers and community scientists who contributed to the eBird database as well as the eBird project team. Funding for this research was provided by the NSF Harnessing the Data Revolution (HDR) program (Award numbers 1940276, 1940176, 1940160, 2023755).

\section*{Data Accessibility}
The datasets and R-code used in the analysis of this study are available on Github at \url{https://github.com/mefeng7/Bird_Outages_MA}. The version of the analysis at the time of publication is archived on Zenodo (link available upon publication).

The compiled power outage data and modeled eBird detection probabilities for the state of MA are also publicly available in Columbia University's International Research Institute for Climate and Society Data Library \cite{blumenthal_iri_2014}. \\
The predicted bird detection probability and outage datasets can be found at:\\ \url{http://iridl.ldeo.columbia.edu/SOURCES/.PRISM/.eBird/.derived/.detectionProbability/} \\ \url{http://iridl.ldeo.columbia.edu/SOURCES/.EOEEA/}

\section*{Conflicts of Interest}
The authors of this manuscript have no conflicts of interest to declare.

\section*{Authors' Contributions}
All authors contributed to the study conception and design; M.-L.E.F and O.O.O collected the data; M.-L.E.F, O.O.O, T.L.J.S., and S.S. analyzed the data; M.-L.E.F and O.O.O led the writing of the manuscript. All authors reviewed, contributed critically, and approved the final manuscript.

\end{spacing}

\bibliographystyle{ieeetr}
\bibliography{ref}

\end{document}


\nolinenumbers

\begin{spacing}{2}   

\section*{Supplement 1: Outage data preprocessing methods}
The initial data preprocessing involved data integrity and data verification procedures to address problems such as typographical errors or mismatch in the location reported for the outage. In the final preprocessing step, corrections were made to account for missing values. 
 The preprocessed data was then used to develop a quantitative risk methodology for measuring outage risk. 
Some of the problems that were addressed during preprocessing includes:
\begin{enumerate}
    \item Missing Data: Missing data in the township feature were filled by locating the street information (where available) on Google Maps \cite{Google}. Out of 138,153 observations, three had townships found in another state and two had no recorded data for town or street. These observations were removed. 
    \item Zero values: Data points with zero customers affected or duration of outages were removed from our analysis. This is because these observations represented no outage occurrence and thus did not assist our analysis.
    \item Typographical errors, extra spaces, and other inconsistencies: A thorough feature inspection was carried out to ensure that the feature elements were consistent throughout. For example, typographical errors in street and township names were addressed using Google Maps \cite{Google}.  
\end{enumerate}

\bibliographystyle{ieeetr}
\bibliography{ref}

\section*{Supplement 2}

\begin{figure}[H]
\centering
\includegraphics[width=13cm]{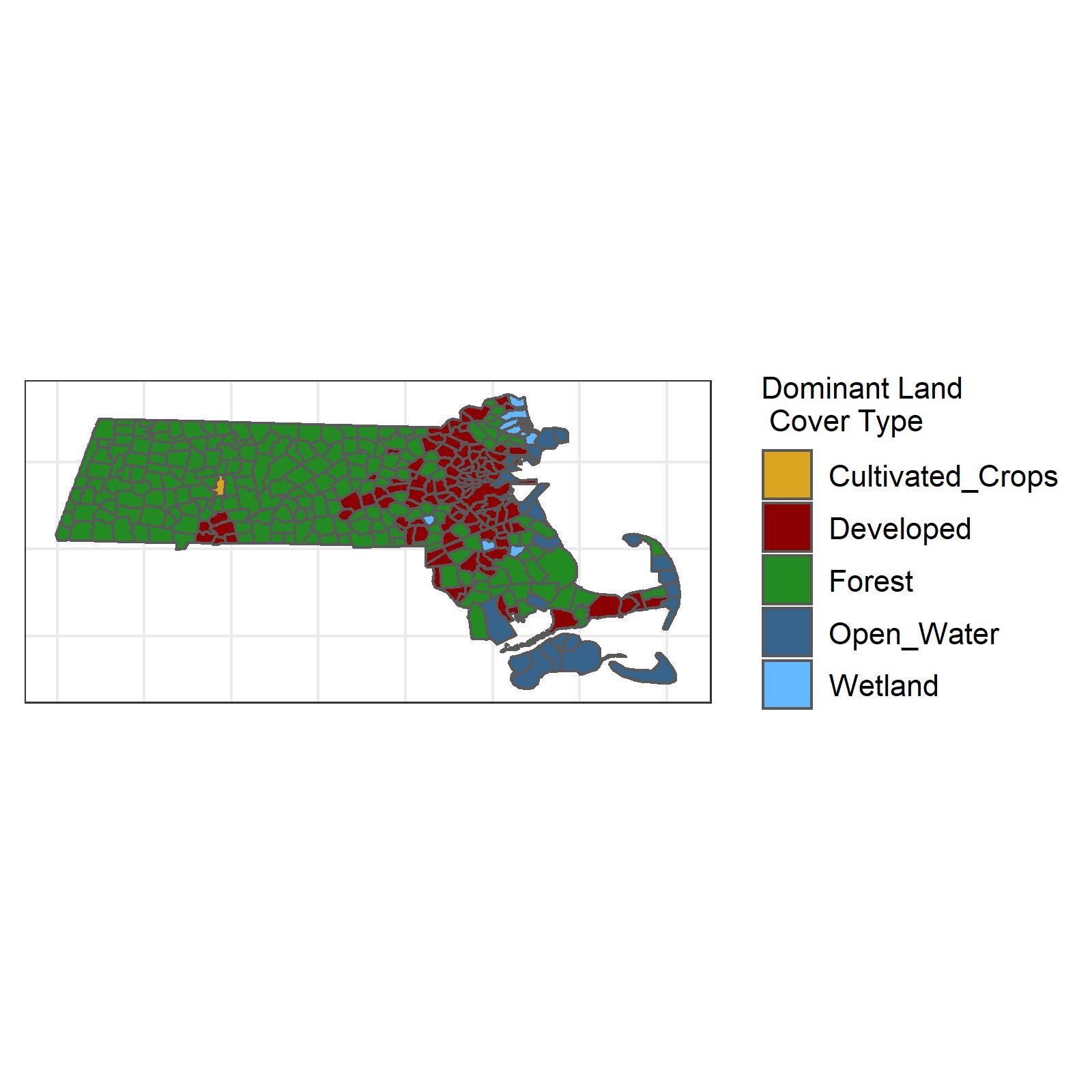}
\caption{Massachusetts townships are mapped by their dominant land cover types in 2018. The proportion of aggregated land cover types from the National Land Cover Database \cite{HOMER2020184} were summarized within each township. The land cover type with the highest proportion was labeled as the township's dominant land cover type.} 
\label{fig:MA_lc}
\end{figure}

\begin{figure}[H]
\centering
\includegraphics[width=13cm]{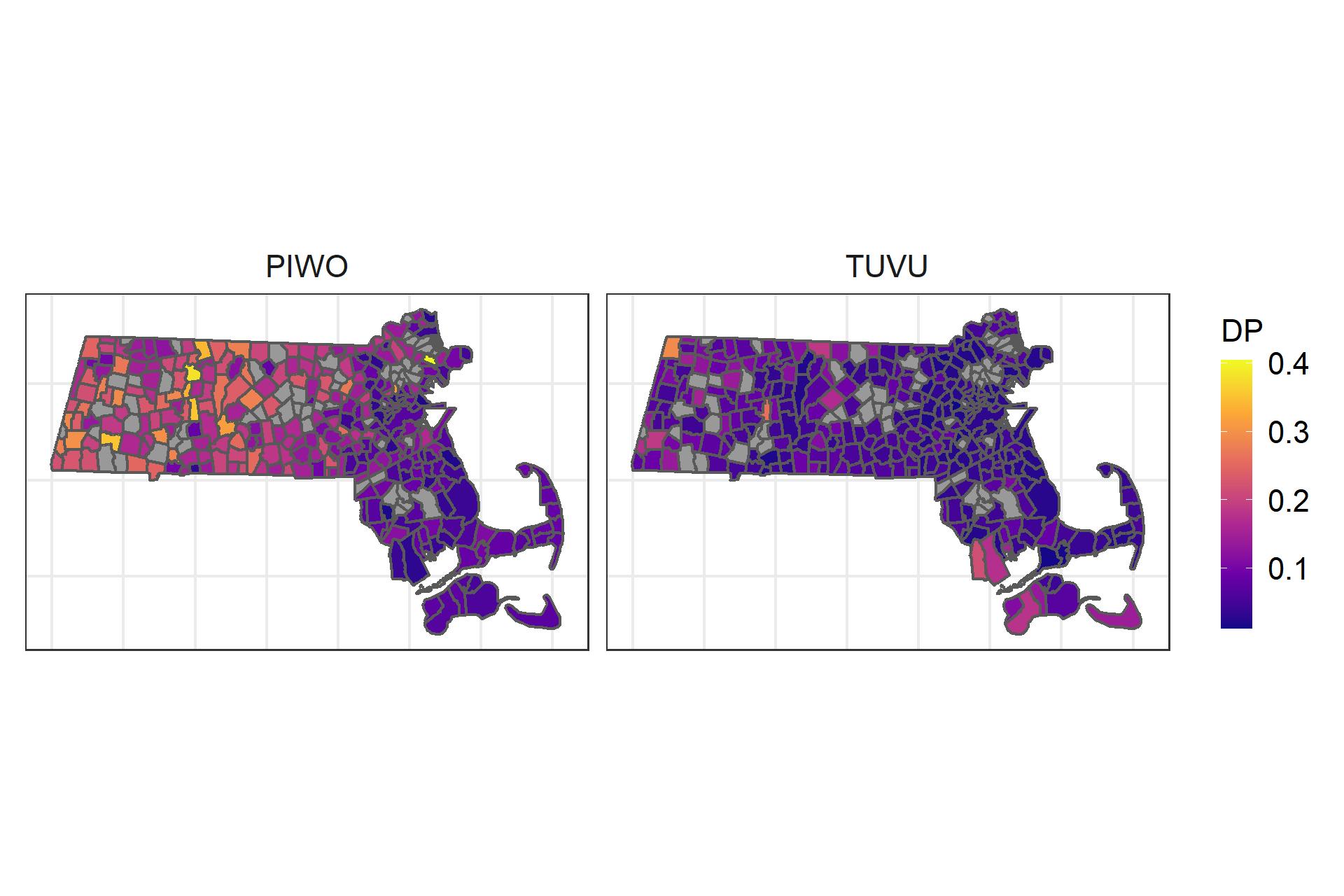}
\caption{Maps of the mean weekly detection probabilities for outage-prone bird species in Massachusetts townships, averaged across weeks from 2013-2018. Townships in grey were not included in our analysis because they did not have available outage data. These species were grouped in the second principal component of detection probabilities and share similar spatial distributions. They occupy rural areas of the state with more continuous, forested habitat.} 
\label{fig:forest_maps}
\end{figure}

\begin{figure}[H]
\centering
\includegraphics[width=13cm]{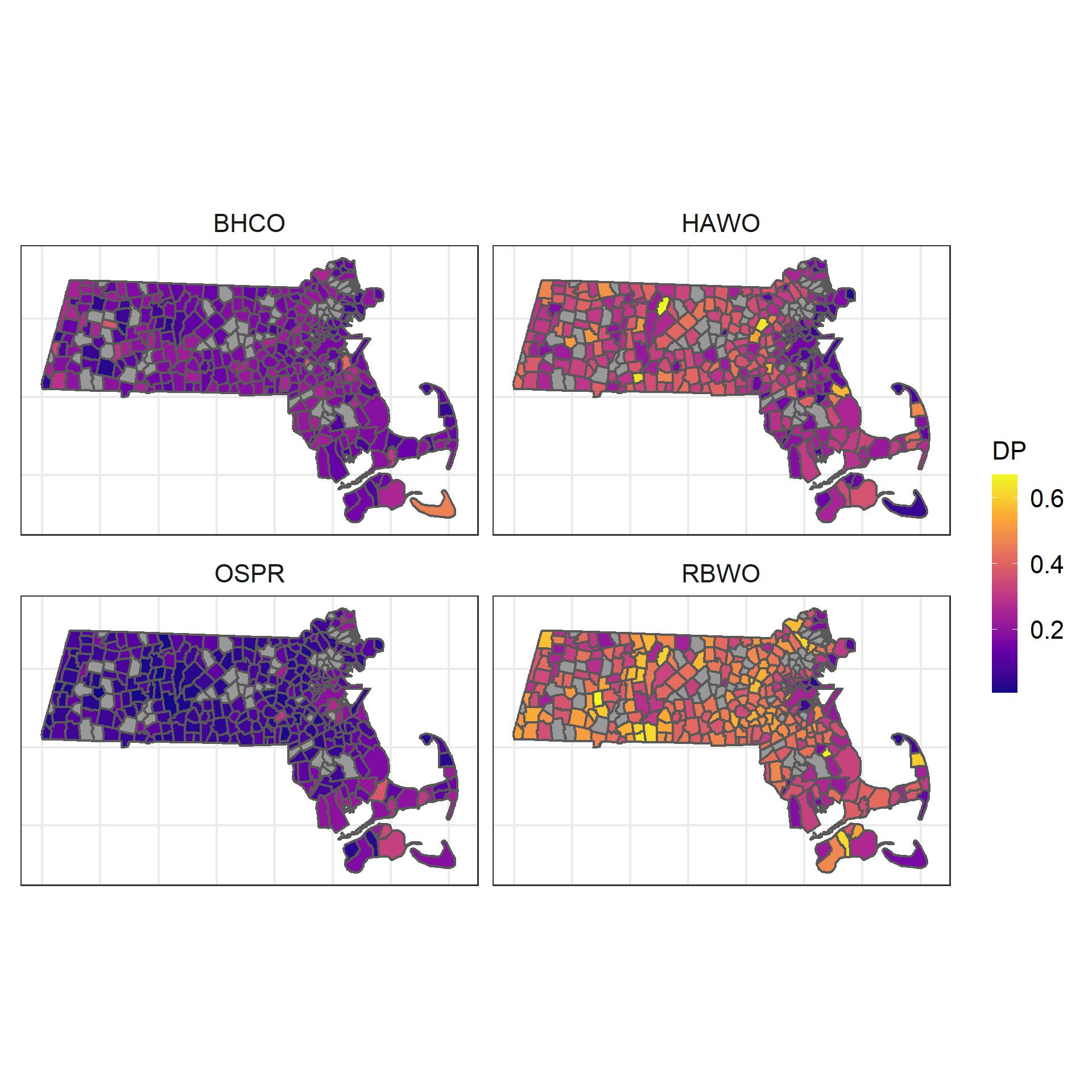}
\caption{Maps of the mean weekly detection probabilities for outage-prone bird species in Massachusetts townships, averaged across weeks from 2013-2018. Townships in grey were not included in our analysis because they did not have available outage data. These species were grouped in the second principal component of detection probabilities and share similar spatial distributions. They do not exhibit a distinct spatial preference towards urbanized or natural habitat.} 
\label{fig:nopattern_maps}
\end{figure}

\begin{figure}[H]
\centering
\includegraphics[width=13cm]{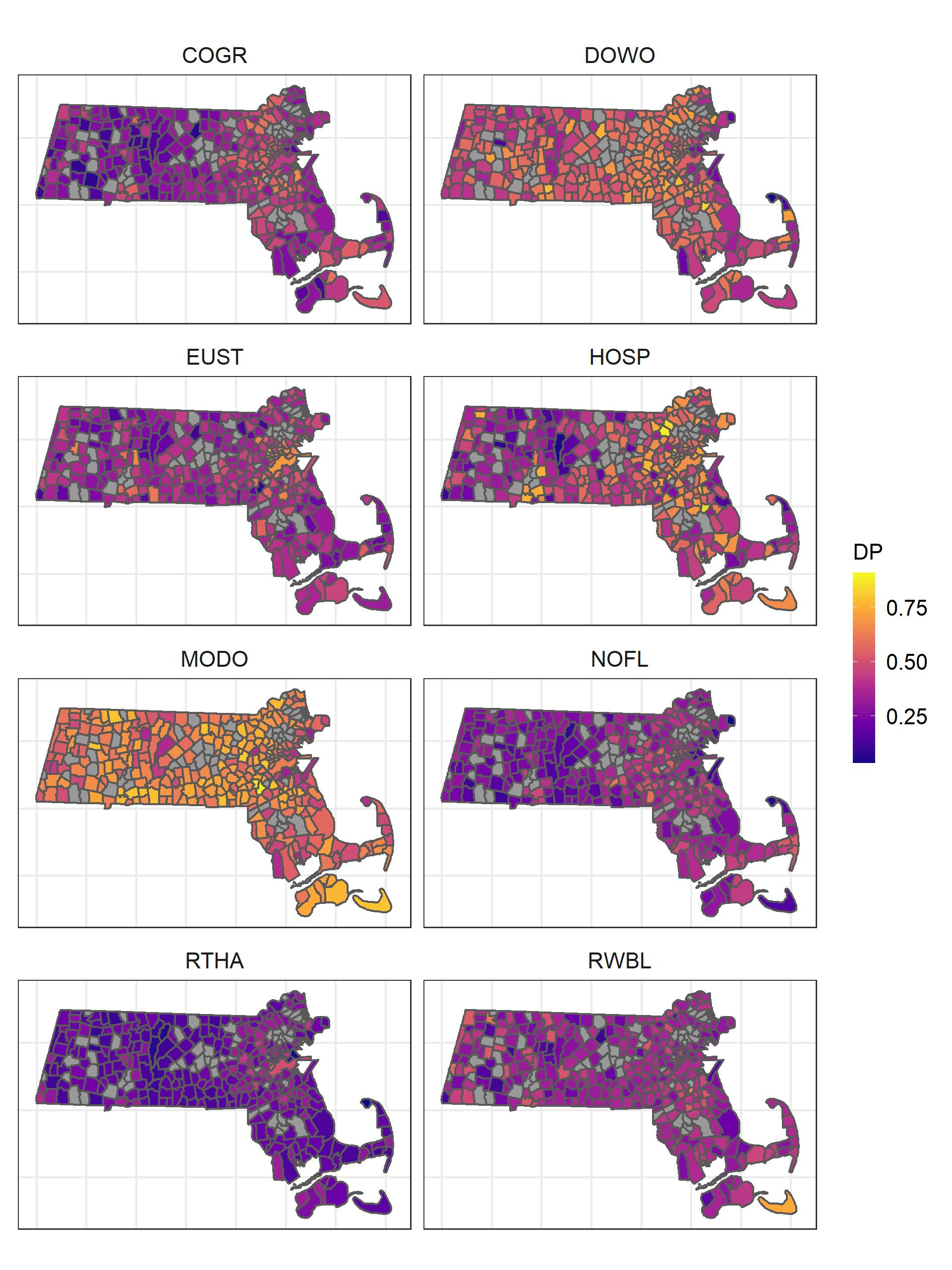}
\caption{Maps of the mean weekly detection probabilities for outage-prone bird species in Massachusetts townships, averaged across weeks from 2013-2018. townships in grey were not included in our analysis because they did not have available outage data. These species were grouped in the second principal component of detection probabilities and share similar spatial distributions. They occupy more urbanized areas surrounding the major cities of Boston and Springfield, MA.} 
\label{fig:urban_maps}
\end{figure}

\bibliographystyle{ieeetr}
\bibliography{ref}

\end{spacing}